\documentclass[a4paper]{jpconf}
\usepackage{iopams}
\usepackage{comment}

\usepackage[a4paper,
            bindingoffset=0.0in,
            left=0.99in,
            right=0.99in,
            top=1.57in,
            bottom=1.063in,
            footskip=0.1in]{geometry}
\usepackage{graphicx,amsbsy,amsmath,epsfig,float,hyperref}

\begin{document}

\title{Collective dynamics and phase transition of active matter
  in presence of orientation adapters}

\author{Sagarika Adhikary and S. B. Santra}

\address{Department of Physics, Indian Institute of Technology Guwahati,
  Guwahati-781039, Assam, India.}
\ead{a.sagarika@alumni.iitg.ac.in, santra@iitg.ac.in}


\begin{abstract}  
  In this work, the orientation adapter, a species of active particles that
  adapt their direction of motion from the other active particles, is
  introduced. The orientation adapters exist besides the usual Vicsek-like
  particles; both are self-driven, however, follow different
  interaction rules. We have studied the dynamics in high speed of the
  particles keeping dissimilar speeds for these different species.
  The effect of orientation adapters on the collective behaviour of the system
  is explored in this model. The orientational order-disorder phase transition
  is mainly studied in such systems. First, for equal density of
  both species, when the adapter speed $v_a=1.2v_0$ and usual particles
  speed $v_0=1.0$, both adapters and the usual particles
form dense travelling bands and move in the same direction. Near the
transition point, such bands appear and
disappear over time, giving rise to the co-existence of two phases.
The adapters and the usual particles
both undergo a discontinuous transition. The nature of the transition
is further confirmed by the existence of hysteresis in the order
parameter under a continuously varying noise field. However, when the
adapter velocity becomes much higher than the usual SPPs $v_a \approx 7v_0$,
the formation of travelling bands disappears from the system, and the
transition becomes continuous. The density ratio is also varied, keeping
the velocities constant, and the phase transition is studied.
For a high adapter velocity with $v_a=10v_0$, the continuous transition
is found with low-density values of the adapters. The critical exponents
related to the continuous transition are also determined.
\end{abstract}

\section{Introduction}
Active or self-propelled particles (SPPs) are known for
self-organisation and collective motion. Fish
schools \cite{becco2006experimental, makris2009critical,
  filella2018model}, flocking birds \cite{ballerini2008interaction,
  cavagna2010scale}, mammalian herds \cite{ginelli2015intermittent},
human crowd \cite{helbing2000simulating, helbing2007dynamics}, swarms
of insects \cite{kelley2013emergent, okubo1974analysis,
  buhl2006disorder, romanczuk2009collective}, bacteria swarms
\cite{shapiro1998thinking,zhang2009swarming, zhang2010collective},
cell clusters \cite{malet2015collective}, actin filaments inside our cells
\cite{schaller2010polar,sanchez2012spontaneous}, and even many
artificial self-driven systems \cite{vicsek2012collective,
  shaebani2020computational,morin2017distortion} are all
examples of active matter which exhibits collective motion. In a
seminal work, Vicsek {\em et al.} proposed a model for studying the
collective motion of SPPs in two dimensions, known as the Vicsek model
(VM). In this model, a large number
of SPPs move together at a constant speed $(v_0)$, and they align
their direction of motion with their neighbours through a short-range
interaction. However, it is subject to an angular noise ($\eta$), present
in the system. For a given density ($\rho_0$), an
orientational order-disorder transition occurs at a critical noise
($\eta_c$). Initially, the nature of this phase
transition in the VM was found to be continuous for low velocity on
small system sizes\cite{nagy2007new, vicsek2012collective}. However,
later it is established through extensive simulations that there
exists a crossover system size $L^*(\rho_0,v_0)$
\cite{chate2008,adhikary2021effect} below which the order of the
transition is continuous, and above which it is discontinuous where
dense bands appear in the system. It needs to be noted that
$L^*(\rho_0,v_0)$ diverges both for low velocities $(v_0<0.05)$ and
low densities $(\rho_0<0.01)$ \cite{chate2008}. The formation of the dense
travelling band near the transition region is fluctuation-driven
and occurs due to the feedback mechanism between local order and local
density \cite{ginelli2016}.

One of the major characteristics of the VM is that all the SPPs have
the same speed, and they all interact with each other locally.
However, in natural systems, the velocities of
particles need not be the same, and also there can be some other type
of species which interact differently. In a recent study of collective
dynamics in a binary mixture of SPPs with widely different velocity
\cite{Adhikary_2022,adhikary2022pattern}, interesting collective patterns
and nontrivial phase transitions are observed. Collective dynamics are studied
with variable speeds of SPPs, which depend on the neighbourhood's polarization
\cite{mishra2012collective,singh2020phase}. Apart from velocities, other
properties of active particles are also varied in many systems. Examples of
such systems include: a mixture of SPPs with different sizes
\cite{dolai2018phase}, a mixture of active Brownian
particles with different diffusion constant
\cite{weber2016binary,demix_sunita}, a mixture of active and passive
particles \cite{stenhammar2015activity, dolai2018phase,
  maloney2020clustering, mccandlish2012spontaneous}, binary active
particles with different alignment interactions \cite{menzel},
an oppositely driven binary mixture of particles
\cite{reichhardt2018laning, Ikeda_2012, bain2017critical}, chiral
active matter \cite{article, liebchen2017collective}, a mixture of
polar and apolar SPPs \cite{sampat2021polar} and many others.
However, the study mixture of species with different interaction properties
in the polar SPPs is a relatively new area of research.
The effect of another type of species, we call them orientation adapters,
on the collective behaviour of usual SPPs, is a crucial aspect to study in the
context of the VM. The orientation adapters are the species that adapt their
direction of motion from other SPPs. We have proposed a model where
adapter SPPs exist besides the usual SPPs in equal or smaller proportions.
The adapter SPPs do not interact among themselves but adopt the velocity
orientation of the usual SPPs through local interactions. However, the
usual SPPs do interact with themselves as well as with the adapters.
The model implements not only different alignment interaction rules,
moreover, their velocities will be dissimilar.
It will be interesting to investigate the effect of adapters on the
order-disorder transition phase transition. What would be
the nature of such a transition? In this study, we explore all these
questions as the adapters induce nontrivial collective behaviour in
the system.

\section{The Model}
A mixture of usual SPPs with adapter SPPs is modelled over a
two-dimensional square box of linear size $L$ with periodic boundary
conditions. The usual SPPs move with a velocity $v_0$, and the adapter
SPPs move with velocity $v_a$. They are taken in equal
proportion. If $N_{0}$ is the number of usual SPPs and $N_{a}$ is the
number of adapters, then $N_{0}=N_{a}=N/2$ (only for the case with
same proportion of the two species), where $N$ is the total
number of SPPs in the system. Initially, the position $\vec{r}_{p,i}$,
$i=1,2,3,\cdots,N/2$ of all the SPPs are randomly distributed over the
space where $p\in\{0,a\}$. The initial orientation $\theta_{p,i}$ of
an SPP is randomly selected in the range $-\pi$ to $\pi$, irrespective
of their type. The usual SPPs interact within a local neighbourhood
$R=1$ and determine their average orientation. Whereas an adapter only
interacts with usual SPPs within the local neighbourhood $R=1$ and
determine their average orientation.

\begin{figure}[t]
\includegraphics[width=14pc]{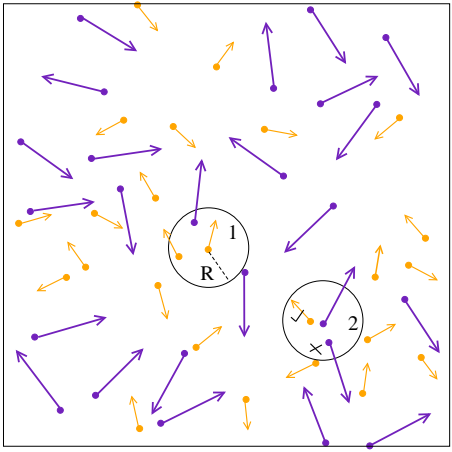}\hspace{3pc}%
\begin{minipage}[b]{20pc}\caption{\label{model}The distribution of
    the SPPs are shown on a
  square box of size $L=10$ where $N_0=N_a=25$. The orange colour represents
  usual SPPs, and the indigo colour represents the adapters. The arrow
  associated with an SPP indicates its orientation, and here $v_a>v_0$.
  A usual SPP present at the centre of circle-$1$ interacts with both 
  SPPs present within the circle of radius $R$. Whereas an adapter
  SPP at the centre of circle-$2$ interacts only with usual
  SPPs present within the circle of radius $R$.}
\end{minipage}
\end{figure}

The distribution of $25$ randomly oriented usual SPPs (in orange) and
$25$ randomly oriented adapters (in indigo) are shown in Fig.\ref{model}.
Longer and shorter arrows show the velocities $v_a$ and $v_0$, respectively
as in this case $v_a>v_0$. The circular region of radius $R$ indicates the
region of interaction.

\hspace{1pc}

The time evolution of the orientation $\theta_{0,i}$ of an usual
SPP is determined by
\begin{equation}
  \label{ch41}
  \theta_{0,i}(t+\Delta t) = {\langle\theta(t)\rangle}_{R\in\{0,a\}}
  + \Delta\theta
\end{equation}
Where, The interaction term $\langle\cdots\rangle_{R\in\{0,a\}}$ for usual SPPs
includes both the usual SPPs ($0$) and the adapters ($a$) within the
radius $R$.  Whereas, time evolution of the orientation $\theta_{a,i}$
of an adapter is determined by
\begin{equation}
  \label{ch42}
  \theta_{a,i}(t+\Delta t) = {\langle\theta(t)\rangle}_{R\in\{0\}} + \Delta\theta
\end{equation}
where, the interaction term $\langle\cdots\rangle_{R\in\{0\}}$ for
adapters only include the usual SPPs within the radius $R$. It should
be noted that the magnitude of the velocity of individual SPPs is ignored
and only the orientations are taken into account in estimating
$\langle\theta(t)\rangle_R$ for both the usual SPPs and the adapters.
Here, $\Delta\theta$ is a random orientation chosen with a uniform
probability from the interval $[-\eta\pi,+\eta\pi]$. The strength of
the angular noise $\eta$ varies from $0$ to $1$ and act as a control
parameter. After averaging, an SPP of type-$p$ ($p\in\{0,a\}$) at the
position $\vec{r}_{p,i}$ is thus moving with a speed $v_p$ in the
direction $\theta_{p,i}$. Knowing the velocity $\vec{v}_{p,i}(t)$ at
every time step, the position of the $i$th SPP $\vec{r}_{p,i}$ is
updated following the forward update rules as given below
\begin{align}
 {\vec{r}}_{0,i}(t+\Delta t) = {\vec{r}}_{0,i}(t) +
 {\vec{v}}_{0,i}(t)\Delta t \label{ch43} \\
 {\vec{r}}_{a,i}(t+\Delta t) = {\vec{r}}_{a,i}(t) +
 {\vec{v}}_{a,i}(t)\Delta t \label{ch44}
\end{align}
where $\Delta t$ is the time between two successive updates, and it is
chosen as $\Delta t=1$. Eqs.\ref{ch41}, \ref{ch42}, \ref{ch43} and
\ref{ch44} are then evolved
with time and dynamical properties of the model are studied, varying
$\eta$ for different velocity ranges.

\section{Phase transition and Finite-size scaling}
Now, we present the results of the orientational order-disorder
phase transition for this model. We analyze the data
for the whole system, considering both types of
SPPs together, as well as the partial systems involving only one type
of SPPs. The order parameter of the transition $\phi$ for the whole
system is defined as
\begin{equation}
  \label{ch4d_opw}
  \phi(\eta,L)= \frac{1}{N}\left|\sum_{p}\sum_{i=1}^{N_p}\frac{{\vec v}_{p,i}}
        {|{\vec v}_{p,i}|}\right|
\end{equation}
where $N$ is the total number of SPPs. The partial
order-parameter of the transition $\phi_p$ for the $p$-type SPPs is
defined as
\begin{equation}
  \label{ch4d_opp}
 \phi_p(\eta,L)= \frac{1}{N_p}\left|\sum_{i=1}^{N_p}\frac{{\vec v}_{p,i}}
      {|{\vec v}_{p,i}|}\right|
\end{equation}
where $N_p$ is the number of $p$-type ($p\in\{0,a\}$) SPPs.

The susceptibility $\chi$ for the whole system and that of the partial
systems $\chi_p$ can be estimated from the fluctuation in their
respective order parameters $\phi$ and $\phi_p$ as
\begin{equation}
  \label{ch4d_susc}
  \chi= L^2 \left[\langle\phi^2\rangle - \langle\phi\rangle^2\right], \ \ \  \chi_p = L^2 \left[\langle\phi_{p}^2\rangle - \langle\phi_{p}\rangle^2\right]
\end{equation}  
where $\langle\phi^n\rangle=\int\phi^nP(\phi)d\phi$,
$\langle\phi_p^n\rangle=\int\phi_p^nP(\phi_p)d\phi_p$, $P(\phi)$ and
$P(\phi_p)$ are the distribution functions of $\phi$ and $\phi_p$
respectively. 
Similarly, the fourth-order Binder cumulant for the whole system and
that of the partial systems are defined as,
\begin{equation}
  \label{ch4d_binder}
U= 1 - \frac{\langle\phi^4\rangle}{3\langle\phi^2\rangle^2}, \ \ \ U_p
= 1 - \frac{\langle\phi_{p}^4\rangle}{3\langle\phi_{p}^2\rangle^2}
\end{equation}
where the higher-order averages are obtained following the definitions
of $\langle\phi^n\rangle$ and $\langle\phi_p^n\rangle$ given above.

If the orientational order-disorder transition is continuous, the finite
size scaling (FSS) relations of the above parameters can be given
following equilibrium critical phenomena
\cite{binder1987theory,christensen2005complexity}, as
\begin{equation}
  \label{d7}
   \phi(\eta,L)=L^{-\beta/\nu}\phi_{0}[\epsilon L^{1/\nu}]
\end{equation} 
where $\epsilon=(\eta-\eta_c)/\eta_c$ the reduced noise, $\beta$ is
the order parameter exponent, $\nu$ is the correlation length exponent
and $\phi_{0}$ is a scaling function. At the criticality
$\eta=\eta_c$, $\phi(\eta_c,L)\sim L^{-\beta/\nu}$. The order
parameter distribution $P_L(\phi)$ for a given system of size $L$ is
defined as
\begin{equation}
  \label{pd}
  P_L(\phi) = L^{\beta/\nu} \widetilde{P}_L\left[\phi L^{\beta/\nu}\right]
\end{equation} 
where $\widetilde{P}_L$ is a scaling function. At the criticality, the
distribution $P_L(\phi)$ is unimodal for a continuous transition. The
FSS form of the susceptibility is given by
\begin{equation}
  \label{d8}
  \chi(\eta,L)= L^{\gamma/\nu}\chi_{0}[\epsilon L^{1/\nu}]
\end{equation}  
where $\chi_0$ is a scaling function, $\gamma/\nu=d-2\beta/\nu$ and
$d$ ($=2$) is the space-dimension. At $\eta=\eta_c$,
$\chi(\eta_c,L)\sim L^{\gamma/\nu}$. The FSS form of the fourth order
Binder cumulant is given by
\begin{equation}
  \label{d9}
  U(\eta,L) = U_{0}[\epsilon L^{1/\nu}]
\end{equation}
where $U_{0}$ is a scaling function. The derivative of $U(\eta,L)$
with respect to $\eta$ follows a scaling relation
\cite{cambui2016critical},
\begin{equation}
  \label{d10}
  U^{\prime}(\eta,L)=L^{1/\nu}\frac{U_{0}^\prime[\epsilon L^{1/\nu}]}{\eta_{c}}
\end{equation}
where the primes on $U$ and $U_0$ denote their derivatives with respect
to $\eta$. For a continuous transition, the cumulant $U$ always remain
positive. At $\eta=\eta_c$, the cumulants of different systems of size
($L$) become independent of $L$ and $U^{\prime}(\eta_c,L)\sim L^{1/\nu}$ at
the transition.

In case the orientational order-disorder transition is discontinuous,
the order parameter exponent $\beta$ should go to zero. As a
consequence, the susceptibility should then scale as $\chi\sim L^d$,
where $d$ is the space dimension. The Binder cumulant $U$ would
exhibit a sharp fall towards a negative value at the transition
point. As the system exhibits the coexistence of two phases, the order
parameter distribution $P(\phi)$ would be a bimodal distribution.

\section{Results with different velocity ratios with fixed
  density ($\rho_{0}=\rho_{a}=0.25$)}
Throughout the simulation, both the SPPs are kept in the same
proportion as $N_{0}=N_{a}=N/2$. The overall particle density
$\rho=N/L^2$ is fixed as $\rho=0.5$ for all the observations.  One
Monte Carlo (MC) time step corresponds to the up-gradation of the position
and orientation of all the particles. Initial $3 \times 10^5$ MC
steps are neglected to achieve the steady state. An ensemble of size
$48\times10^5$ is taken for statistical averages ($2\times 10^5$
samples at different times for
$24$ different initial configurations). The results are
shown with the velocity of the adapters as $v_a> v_0$ for a fixed
velocity of the usual SPPs $v_0=1.0$.

\begin{figure}[t]
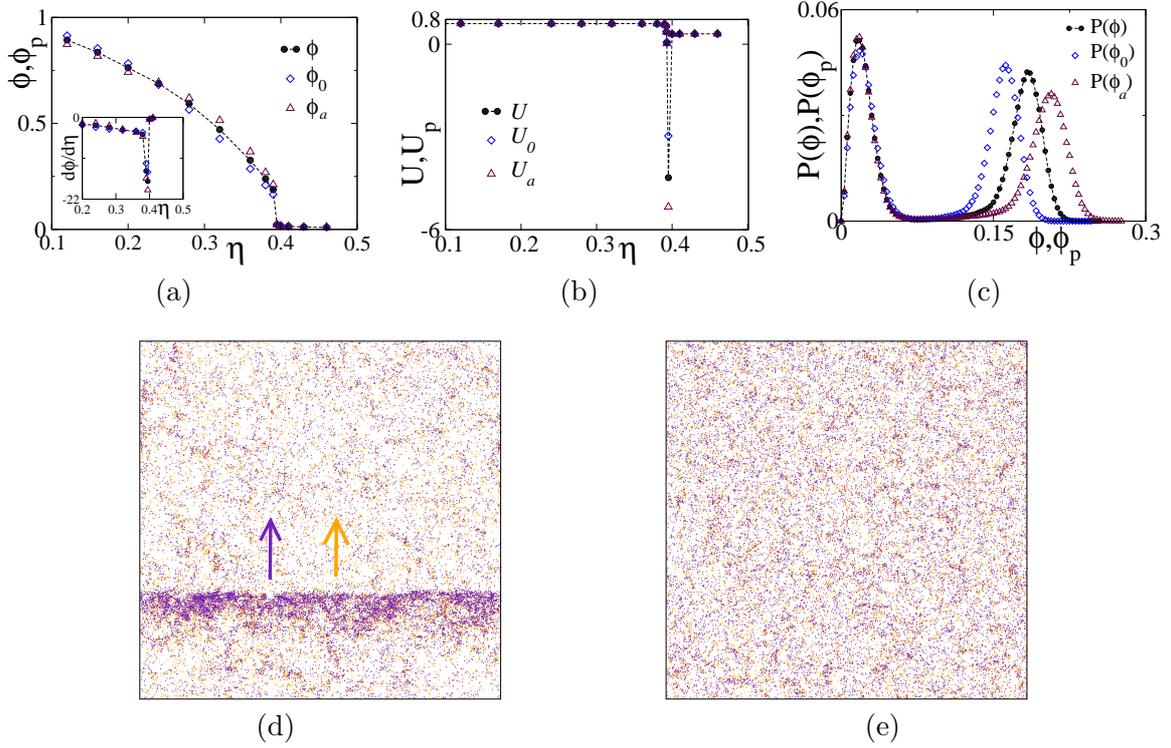

    \centerline{
      \hfill\psfig{file=plots/op_1.2_256.eps,width=0.30\textwidth}
  \hfill\psfig{file=plots/bin_1.2_256.eps,width=0.30\textwidth}
    \hfill\psfig{file=plots/pdf_1.2_256.eps,width=0.30\textwidth}
   \hfill }
    \centerline{\hfill(a) \hfill\hfill(b) \hfill\hfill(c) \hfill}
        \vspace{1.0em}
\centerline{
  \hfill\psfig{file=plots/confi_12.eps,width=0.3\textwidth}
        \hfill\psfig{file=plots/confi_12_2.eps,width=0.3\textwidth}
        \hfill }
\centerline{\hfill (d) \hfill\hfill (e) \hfill}
\caption{\label{op_p2} For $v_{0}=1.0$, $v_{a}=1.2v_0$: (a) Plot of
  $\phi$ and $\phi_p$ versus $\eta$.
  Derivatives of $\phi$ and $\phi_p$ with respect to $\eta$ are shown in
  the inset. (b) Plot of $U$ and $U_p$ versus $\eta$. (c) Plot of $P(\phi)$
  and $P(\phi_p)$ at $\eta=\eta_c$. System
    morphology for $\eta\approx\eta_c$ with (d) ordered phase (density band)
    and (e) disordered phase. Orange colour represents the usual SPPs, and
    indigo colour represents the adapters. System size is $L=256$.}
\end{figure}

\subsection{Results with $v_{a}=1.2 v_0$ and $v_0=1.0$:}
First, we present data for $\phi$ (and $\phi_p$), $U$ (and $U_p$), 
and $P(\phi)$ (and $P(\phi_p)$) respectively in Fig.\ref{op_p2}(a), (b) and
(c) on a system of size $L=256$ for $v_a=1.2 v_0$, $v_0=1.0$. In this
case, the velocity $v_a$ of adapters is almost similar to the usual SPPs.
The order parameter $\phi$ of the whole system and $\phi_p$, that
of the partial systems, are plotted against $\eta$ in Fig.\ref{op_p2}(a).
There are jumps in the values of $\phi$ and
$\phi_p$ near the transition. The values of $\phi$, $\phi_0$
and $\phi_a$ are almost similar for a given $\eta$ in this
case. The derivatives of $\phi$ and $\phi_p$
with respect to $\eta$ are plotted in the inset of
Fig.\ref{op_p2}(a) and sharp minima are observed at
$\eta_c\approx 0.39$ (same for the whole and the partial systems). The
Binder cumulants $U$ and $U_p$ versus $\eta$ plots are shown in
Fig.\ref{op_p2}(b). The cumulant for the usual SPPs ($U_0$),
for adapters ($U_a$) and for the whole system ($U$) all show
a sharp negative dip at the transition. It implies the discontinuous
transition for the whole as well as for the partial systems. Then, the
distributions of order parameters $P(\phi)$ and $P(\phi_p)$ at
$\eta=\eta_c$ are shown in Fig.\ref{op_p2}(c). In this case, all
three distributions $P(\phi)$, $P(\phi_0)$ and $P(\phi_a)$ exhibit
bimodal distributions, which implies the two-phase co-existence in the
system.

The system morphology is shown in the ordered phase with $\eta=0.36$ and
in the disordered phase with $\eta=0.40$, respectively, in Fig.\ref{op_p2}(d)
and (e). Near the transition, in the ordered phase (Fig.\ref{op_p2}(d)), both
the adapters and usual SPPs are observed in the density band, and they
move together in a particular direction (shown by arrows).
This is expected as the adapters follow the usual SPPs; they move
in the density band's direction. Moreover, their
velocity is similar to the usual SPPs; they also form a dense
band pattern near the usual SPPs. Whereas, at the disordered phase, no such
dense band forms, and all the particles moves randomly (Fig.\ref{op_p2}(e)).
In the case of discontinuous transition, at the transition point, dense
travelling bands of SPPs periodically form and disappear with time, resulting
in the coexistence of two phases in the system. The situation will be different
for a much higher velocity of the adapters.

\begin{figure}[t]
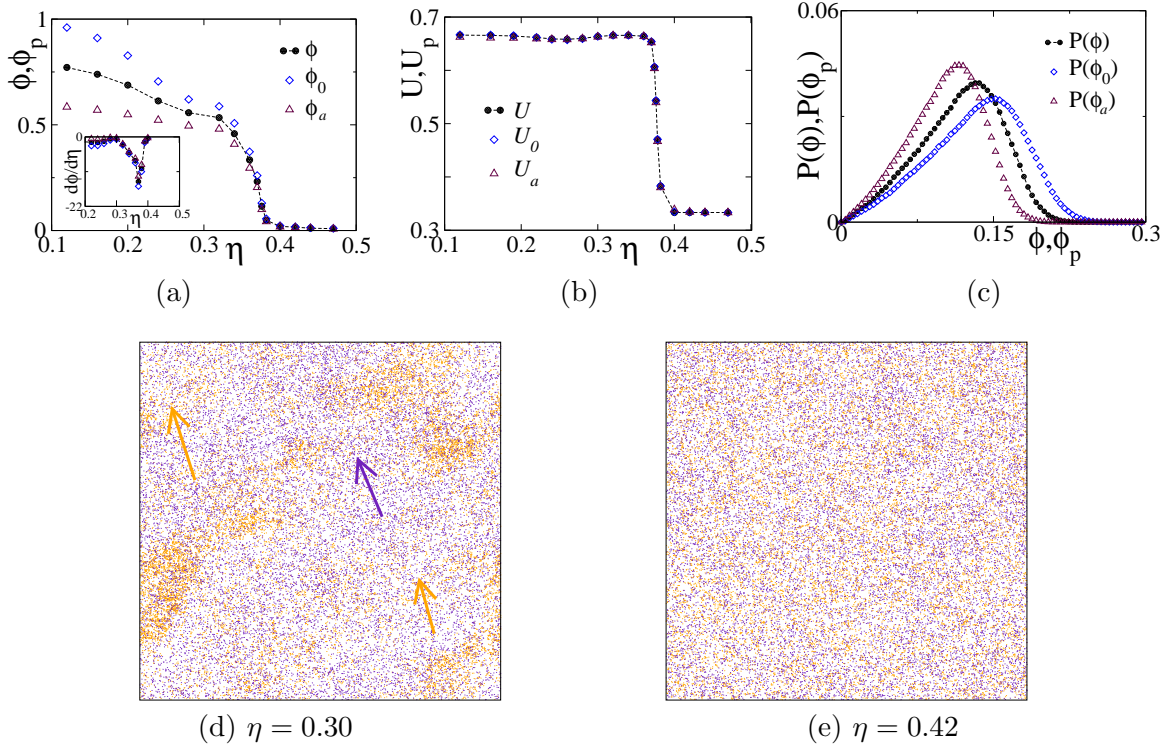

\centerline{
  \hfill\psfig{file=plots/op_7_256.eps,width=0.30\textwidth}
  \hfill\psfig{file=plots/bin_7_256.eps,width=0.30\textwidth}
    \hfill\psfig{file=plots/pdf_7_256.eps,width=0.30\textwidth}
    \hfill }
\centerline{\hfill(a) \hfill\hfill(b) \hfill\hfill(c) \hfill}
    \vspace{1.0em}
\centerline{
  \hfill\psfig{file=plots/confi_7_order.eps,width=0.3\textwidth}
  \hfill\psfig{file=plots/confi_7_disorder.eps,width=0.3\textwidth}
        \hfill }
\centerline{\hfill (d) $\eta=0.30$ \hfill\hfill (e) $\eta=0.42$ \hfill}
  \caption{\label{op_p22} For $v_{0}=1.0$, $v_{a}=7.0v_0$:
  (a) Plot of $\phi$ and $\phi_p$ versus $\eta$. Derivatives of
  $\phi$ and $\phi_p$ with respect to $\eta$ are shown in the
  inset. (b) Plot of $U$ and $U_p$ versus $\eta$. (c) Plot of $P(\phi)$
  and $P(\phi_p)$ at $\eta=\eta_c$. System
    morphology for (d) ordered phase $\eta=0.30$ and (e) disordered
    phase $\eta=0.42$. Orange colour represents the usual SPPs
    , and the indigo colour represents the adapters. System size is $L=256$.}
\end{figure}

\subsection{Results with $v_{a}=7.0 v_0$ and $v_0=1.0$:}
Next, the situation is going to be drastic if the velocity $v_a$ is
much higher than $v_0$. We present the data
$\phi$ (and $\phi_p$), $U$ (and $U_p$), and $P(\phi)$ (and $P(\phi_p)$)
respectively on a system of size $L=256$ for $v_a=7.0 v_0$, $v_0=1.0$
in Fig.\ref{op_p22}(a), (b) and (c), respectively. 
The order parameters $\phi$ and $\phi_p$, are plotted against
$\eta$ in Fig.\ref{op_p22}(a). The order of the usual SPPs ($\phi_0$)
is higher than the order of adapters ($\phi_a$) in the low $\eta$
region in this case. However, near the transition, their values
are similar. The values of $\phi$ and $\phi_p$
are decreasing smoothly to zero as $\eta$ increases. The derivatives
of $\phi$ and $\phi_p$ with respect to $\eta$ are plotted in the inset
of Fig.\ref{op_p22}(a) and minima of the plots at the transition
noise $\eta_c\approx 0.37$ (same for the whole and partial systems)
are observed. In Fig.\ref{op_p22}(b), $U$ and $U_p$ are plotted against
$\eta$. Both $U$ and $U_p$ remain positive over the whole range of
$\eta$. In Fig.\ref{op_p22}(c), the distribution of order
parameters $P(\phi)$ and $P(\phi_p)$ are plotted at
$\eta=\eta_{c}$. All the distributions are unimodal.
The positive Binder cumulants and unimodal distributions of order
parameters indicate a continuous transition in the whole system as well
as in the partial systems for the case of $v_a=7.0v_0$ and $v_0=1.0$.

The system morphologies for the ordered phase with $\eta=0.30$
and for disordered phase with $\eta=0.42$ are shown in
Fig.\ref{op_p22}(d) and (e) respectively for the case of
$v_a=7.0v_0$ and $v_0=1.0$. Below the transition, at $\eta=0.30$
shown in Fig.\ref{op_p22}(d), less dense large clusters of usual
SPPs are observed with small clusters of adapters, and they are directed
in the same direction (shown by arrow). Whereas, above the transition,
at $\eta=0.42$ shown in Fig.\ref{confi_p2_7}(e), they move randomly.
In this case, adapters move with much higher velocity and cannot flock with
the usual SPPs. Moreover, they adopt the alignment information
from usual SPPs, which are distant.
It seems that as the usual SPPs interact with
the adapters, randomness enters into the alignment of the usual
SPPs. Hence, the correlation between the usual SPPs gets destroyed
to form a dense band structure. 

\noindent{\bf Hysteresis study:}
It is well known that the hysteresis phenomena usually accompany the
first-order phase transition \cite{chate2008, durve2016first, pre2019}
which occurs near the transition. New simulations are carried out to
measure the instantaneous order parameter ($\phi_t$) by either gradually
increasing or decreasing the angular noise $\eta$ with a fixed ramp
rate, where each previous state will be implemented as the initial
state of the next simulation process with new $\eta$. The ramp rate
used here is $1.27 \times 10^{-6}$ in radians/unit time. Each hysteresis
loop is obtained by averaging over $800$ independent realizations.
On ramping the angular noise parameter $\eta$ at the same ramping
rate up and down through the transition point,
a hysteresis loop is formed in the case of the first-order phase
transition and the loop area imply phase coexistence.

\begin{figure}[t]
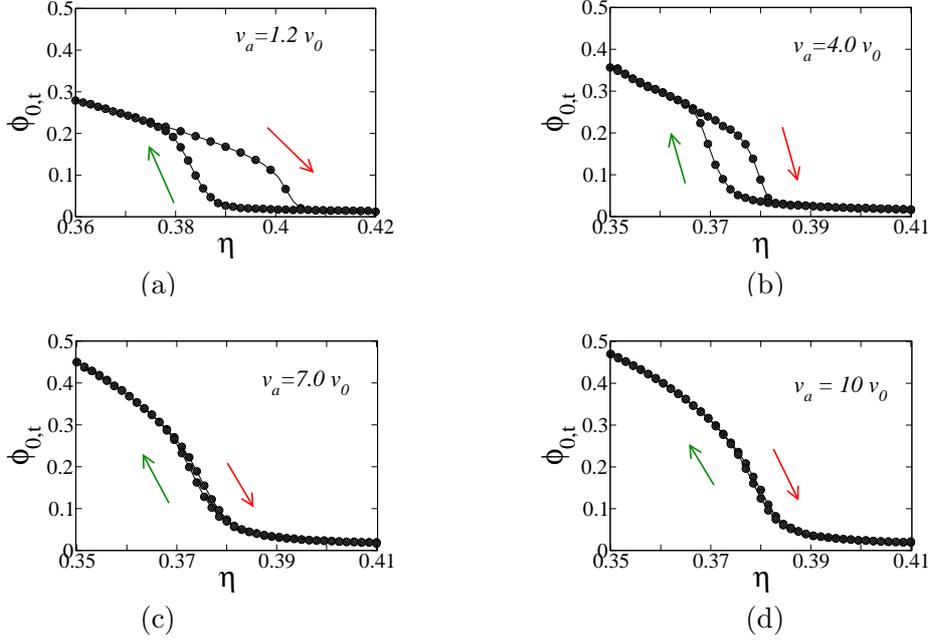

\centerline{
  \hfill\psfig{file=plots/hyst_12.eps,width=0.32\textwidth}
  \hfill\psfig{file=plots/hyst_40.eps,width=0.32\textwidth}
  \hfill }
\centerline{\hfill (a) \hfill\hfill (b) \hfill}
     \vspace{1.0 em}
\centerline{
  \hfill\psfig{file=plots/hyst_70.eps,width=0.32\textwidth}
  \hfill\psfig{file=plots/hyst_100.eps,width=0.32\textwidth}  
  \hfill }
\centerline{\hfill (c) \hfill\hfill (d) \hfill}
  \caption{\label{p2_hyst} For $v_a > v_0$: Hysteresis plot for
    the usual SPPs with different values of adapter velocity as
    (a) $v_a=1.2 v_0$,
    (b) $v_a=4.0 v_0$, (c) $v_a=7.0 v_0$ and (d) $v_a=10 v_0$
    with the fixed $v_0=1.0$.
    Results obtained by implementing ramp-up or forward
    (red arrow) and ramp-down or backward (green arrow) simulation
    schemes, respectively. }
\end{figure}

Simulations are carried out to measure the
instantaneous order parameter of the usual SPPs ($\phi_{0,t}$) by
either gradually increasing or decreasing $\eta$ with a fixed ramp rate
as discussed. The simulation results of hysteresis with different
values of adapter's velocity as $v_{a}=1.2 v_0$, $4.0v_0$, $7.0v_0$ and
$10v_0$ for a fixed $v_0=1.0$, are
presented in Fig.\ref{p2_hyst}(a), (b), (c) and (d), respectively.
Observations show that when $v_a$ is relatively less as
$v_a=1.2v_0$ with $v_0=1.0$, there arise an abrupt
jump in the $\phi_t$ and the positions of the jump by the
implementation of the forward and backward simulations schemes are
different. So the emergence of a hysteresis loop indicates 
the irreversibility of transition \cite{chate2008}, shown
in Fig.\ref{p2_hyst}(a).
Hysteresis loop still exists for a moderately high value
of $v_a=4.0v_0$ shown in Fig.\ref{p2_hyst}(b); however, it
is less prominent than the $v_a=1.2v_0$ case. 
Then, for a large enough value of $v_{a}=7.0 v_0$ and $v_a=10v_0$
the change of the order parameter versus $\eta$ becomes
reversible, which indicates continuous phase transition, as shown in
Fig.\ref{p2_hyst}(c) and (d). It needs to be noted that similar behaviour
of hysteresis plots for the different values of $v_a$
is also observed for the whole system.

\section{Results with different density ratio with $v_{a}=10 v_0$
  ($v_0=1.0$)}
\begin{figure}[t]
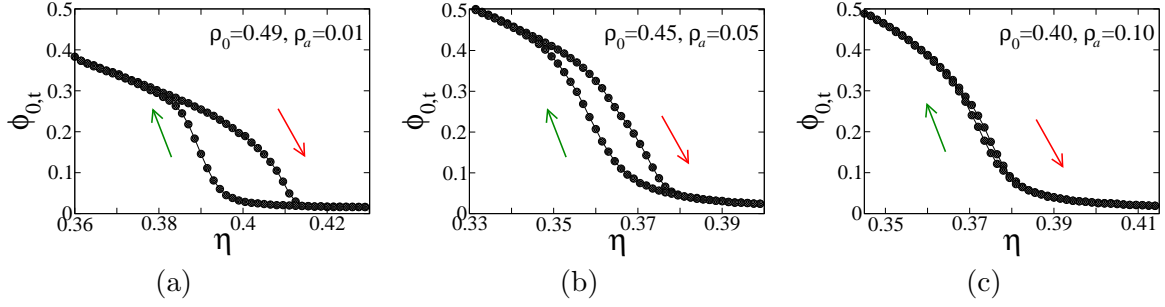

\centerline{
  \hfill\psfig{file=plots/hyst_49_01.eps,width=0.3\textwidth}
  \hfill\psfig{file=plots/hyst_45_05.eps,width=0.3\textwidth}
\hfill\psfig{file=plots/hyst_40_10.eps,width=0.3\textwidth}
  \hfill }
\centerline{\hfill (a) \hfill\hfill (b) \hfill\hfill (c)
  \hfill}
  \caption{\label{p3_hyst} For $v_a=10v_0$: Hysteresis plot for
    the usual SPPs with
    (a) $\rho_0=0.49$, $\rho_a=0.01$,
    (b) $\rho_0=0.45$, $\rho_a=0.05$ and (c) $\rho_0=0.40$, $\rho_a=0.10$.
    Results obtained by implementing
    ramp-up or forward (red arrow) and ramp-down or backward (green arrow)
    simulation schemes, respectively.}
\end{figure}

As it is seen, the high velocity of the adapters induces a continuous
transition when they are present in equal densities with the usual
SPPs. For the fixed velocities of the two types $v_{a}=10 v_0$ and $v_0=1.0$,
we now see the effect of different density ratios on the collective
behaviour. If $\rho_a=0$, the model is equivalent to the VM ($\rho_0=0.5$).
Then a discontinuous order-disorder is expected to occur
in the system for such $v_0$, $\rho_0$. If the $rho_a$ is increased
and $rho_0$ is decreased, keeping the total density fixed at $0.5$, how it
affects the transition is interesting to observe. For that, hysteresis is
studied for different values of $\rho_a$ and $\rho_o$.
Simulations are carried out to measure the
instantaneous order parameter of the usual SPPs ($\phi_{0,t}$) by
either gradually increasing or decreasing $\eta$ with a fixed ramp rate
as discussed earlier. The simulation results of hysteresis with different
density ratios are presented in Fig.\ref{p3_hyst}, with the fixed velocities
$v_{a}=10v_0$ and $v_0=1.0$. 
Observations show that when $\rho_a=0.01$ is relatively less, there arise
an abrupt jump in the $\phi_{0,t}$ and the positions of the jump by the
implementation of the forward and backward simulations schemes are
different, shown in Fig.\ref{p3_hyst}(a).
Hysteresis loop still exists for an increase in the number
of adapters as $\rho_a=0.05$ shown in Fig.\ref{p3_hyst}(b); however, it
is less prominent than the $\rho_a=0.01$ case. 
Then, for a large enough value of $\rho_{a}=0.40$,
hysteresis disappears, and the change of the order parameter versus
$\eta$ becomes reversible, which indicates continuous phase transition,
shown in Fig.\ref{p3_hyst}(c). It needs to be noted that similar behaviour
of hysteresis plots for the different values of $\rho_a$
is also observed for the whole system.

\subsection{Finite-size scaling analysis:}
\begin{figure}[t]
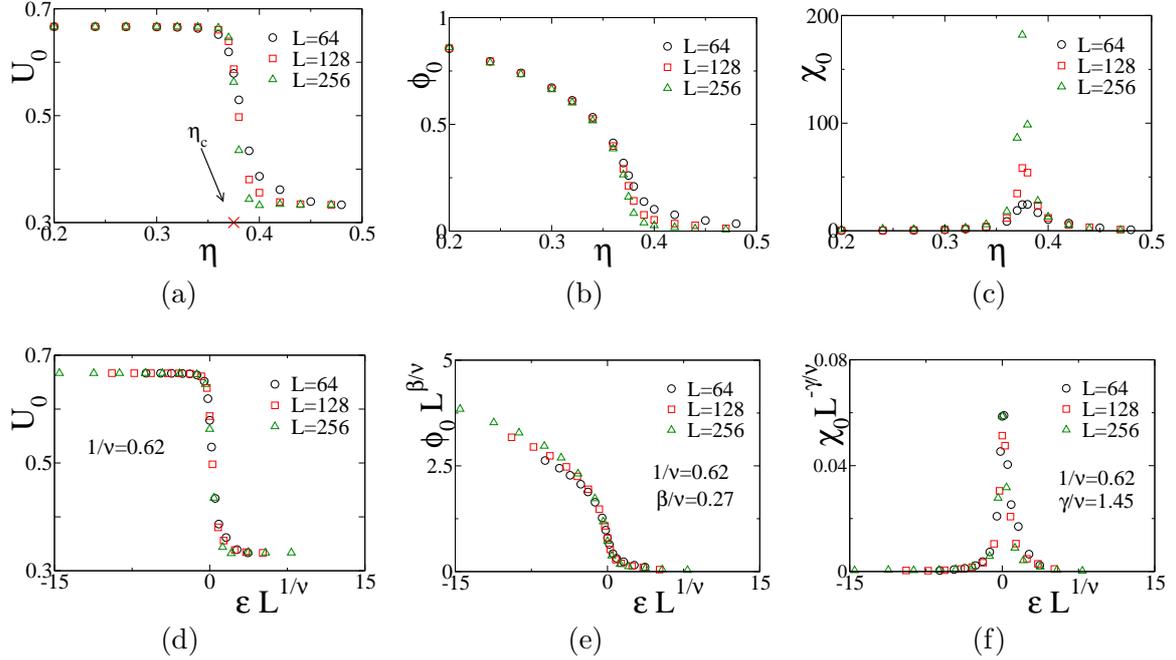

\centerline{
  \hfill\psfig{file=plots/binder_L.eps,width=0.3\textwidth}
  \hfill\psfig{file=plots/op_L.eps,width=0.3\textwidth}
   \hfill\psfig{file=plots/sus_L.eps,width=0.3\textwidth}
   \hfill }
\centerline{\hfill (a) \hfill\hfill (b)  \hfill\hfill (c)
  \hfill}
     \vspace{1.2 em}
\centerline{ 
  \hfill\psfig{file=plots/fss_binder.eps,width=0.3\textwidth}
  \hfill\psfig{file=plots/fss_op.eps,width=0.3\textwidth}
   \hfill\psfig{file=plots/fss_sus.eps,width=0.3\textwidth}
  \hfill }
\centerline{\hfill (d) \hfill\hfill (e)  \hfill\hfill (f)
  \hfill}
\caption{\label{fss} $v_a=10v_0$, $\rho_0=0.40$, $\rho_a=0.10$:
  (a) Plot of $U_0$, (b) $\phi_0$ and (c) $\chi_0$
  versus $\eta$ for $L=64$, $128$ and $256$. In plot (a),
  the cross on the $\eta$-axis indicates $\eta_c$. (d) Plot of $U_0$
  versus the scaled noise $\epsilon L^{1/\nu}$. (e) Plot of $\phi_0 L^{\beta/\nu}$
   against $\epsilon L^{1/\nu}$. (f) Plot of $\chi_0 L^{-\gamma/\nu}$
   against $\epsilon L^{1/\nu}$. The values of the exponents are taken
   as $\beta/\nu=0.27$, $\gamma/\nu=1.45$ and $1/\nu=0.62$.}
\end{figure}

The critical exponents are extracted for the usual particles with
$v_a=10v_0$ and $\rho_0=0.40$, $\rho_a=0.10$ performing FSS analysis.
Binder cumulant $U_0$, order parameter $\phi_0$ and susceptibility $\chi_0$
are plotted against the angular noise $\eta$ for three different systems
of sizes $L=64$, $128$ and $256$ in Fig.\ref{fss}(a), (b) and (c)
respectively. The plots of $U_0$ versus $\eta$ 
for different $L$ intersect at $\eta_c\approx 0.37$, the $L$
independent critical point as expected in a continuous transition. It
is marked by a cross on the $\eta$-axis. A rough estimate of
the exponents $1/\nu$, $\beta/\nu$ and $\gamma/\nu$ are obtained from
the scaling relations $U_0^{\prime}(\eta_c,L)\sim L^{1/\nu}$,
$\phi_0(\eta_c,L)\sim L^{-\beta/\nu}$ and $\chi_0(\eta_c,L)\sim
L^{\gamma/\nu}$ at the criticality. The best possible FSS forms of the
scaled parameters against the scaled noise $\epsilon L^{1/\nu}$ are
obtained, tuning these exponents further. $U_0$, $\phi_0 L^{\beta/\nu}$,
and $\chi_0 L^{-\gamma/\nu}$ are plotted against $\epsilon L^{1/\nu}$ in
Fig.\ref{fss}(d), (e) and (f), respectively. A reasonable
collapse of data in all three cases is obtained by taking $1/\nu=0.62$,
$\beta/\nu=0.27$ and $\gamma/\nu=1.45$ at $\eta_c=0.37$. The critical
exponents satisfy the scaling relation $\gamma/\nu+2\beta/\nu=2$
within error bars. The exponents are similar to the exponents
for the VM \cite{baglietto2008finite} with $v_0=0.1$ and density
$\rho=1/8$ to $3/4$.

\section{Summary and Discussion}
In this work, the collective motion of a mixture of usual SPPs and
orientation adapters is modelled. The two species have different velocities. 
The adapters have higher velocity than the usual SPPs,
while the velocity of the usual SPPs is fixed at a high velocity
$v_0=1.0$. For the same density of the two species, with the adapter velocity
$v_a=1.2v_0$, both adapters and the usual SPPs form dense travelling bands
in the system. Both the travelling bands move in the same direction in
accordance with the orientation rule. Near the transition point, such
bands appear and disappear over time, giving rise to co-existence
of two phases. The adapters and the usual SPPs both undergo a
discontinuous transition characterized by a negative dip in the Binder
cumulant and bimodal distribution of order parameters. The nature of
the transition is further confirmed by the existence of hysteresis in the
order parameter under a continuously varying noise field. This is
quite expected in the context of the VM.
However, more dramatic effects are revealed as the adapter velocity
becomes much higher than the usual SPPs. Surprisingly, the
formation of travelling bands disappears from the system. In the ordered
phase, the flocks of usual SPPs form directed clusters, and the
adapters form relatively smaller directed clusters. All these
clusters move in the same direction as expected. In the disordered phase
($\eta>\eta_c$), these clusters melt into smaller clusters and move
randomly. Consequently, continuous transitions occur for both the
adapters and the usual SPPs, even at such high velocities. The
continuous transitions are characterized by positive Binder cumulant
and unimodal distribution of the order parameter. The hysteresis loops
also disappear for these systems. Such continuous transition is also observed
even for a smaller fraction of adapters ($\rho_a=0.10$, $\rho_0=0.40$)
with high velocity $v_a=10v_0$. Furthermore, the values of critical exponents
related to the continuous transitions are determined. In this system,
the alignment of an adapter obtained
from local interaction may be very different from the alignment of
SPPs at a distant point where the adapter moves after position update
due to its high velocity. Such misalignment in orientations between
the adapters and the SPPs introduces extra fluctuations in the
system. Such fluctuations grow predominantly in the transition region,
and all long-range correlations get destroyed. The continuous nature
of the transition is essentially a manifestation of such a smooth crossover
from a correlated system to an uncorrelated system.

\noindent{\bf Acknowledgement:} The computational facility HPC Newton
and Param-Ishan provided by the Department of Physics, Indian Institute of
Technology Guwahati is gratefully acknowledged.

\section*{References}

\bibliographystyle{iopart-num}
\bibliography{ref2}

\end{document}